# ACHIEVING CONSENSUS WITH INDIVIDUAL CENTRALITY APPROACH

Vitri Tundjungsari[1], Jazi Eko Istiyanto, Edi Winarko, Retantyo Wardoyo

Department of Computer Science and Electronics
Gadjah Mada University
Jl. Sekip Utara
Yogyakarta, Indonesia

Vibara11@gmail.com

## ABSTRACT

*This paper proposes a new consensus model in participatory decision making. The model employs advice centrality approach by electing a leader and recommender named as Supra Decision Maker (SDM). A SDM has a role as a decision bench-marker to other decision makers in evaluating each alternative with respect to given criteria. The weighting value for each alternative can be obtained by considering consensus level and preferences' distances between SDM and other Decision Makers. A social function using Social Judgment Scheme (SJS) concept is employed when a decision does not achieve the required consensus level. A simple example is presented here to illustrate our model.*

## KEYWORDS

*Consensus, Group decision making, Centrality, Supra Decision Maker, Social Judgment Scheme*

## 1. INTRODUCTION

Decision making theory recently has been used to support and facilitate citizen as stakeholder in societal decision making and deliberative democracy as mentioned in [7], [10], [15]. Citizen participation and engagement changes the perspective that citizens not only act as the receiver of decision but also actively perform in making the decision, which is also called as participatory approach. However, this approach requires more complex methods and procedures compare to individual decision or a group decision making involving expertise only. Group decisions are more complex compared to single decision making, since a number of contradicting factors are involved such as individuals' personal opinions, goals and stakes resulting in a social procedure, where negotiation and strategy plays a critical role. Despite the inherent complexity within a participatory group decision making, members are able to express personal opinions and suggest solutions from a personal perspective increasing thus decision outcome efficiency. Several methods have been applied to support participatory approach, such as: consensus ([6], [22]), negotiation [9], and voting [15]. Each method offers its benefit, such as: simple and clear procedure, better efficiency of decisions, dispute minimization. However, participatory decision making in social environments raises some issues, such as: conflicting individual goals, not sufficient knowledge, validity of information and individuals' motivation.

Despite the difficulties, we argue that consensus-based decision making is suitable approach for supporting participatory approach. In this paper we present our proposed consensus method, as a modification to [6] which utilizes consensus level and social function in multi-criteria environment to support participatory decision making. In section 2, we present an overview of several methods applied in decision making theory. Section 3 of this paper discusses the proposed consensus model. We also illustrate our model in section 4, in order to give a clear example in implementation. Finally, section 5 presents conclusion to summarize the paper.

## 2. APPROACHES IN GROUP DECISION MAKING

Authors in [14] state three basic approaches towards group decision-making. i.e.:

1. *Game theory*. This approach implies a conflict or competition between the decision makers. Game theory can be defined as the study of mathematical models of conflict and cooperation between intelligent and rational decision makers. Modern game theory gained prominence after the work of Von Neumann in 1928. Game theory became an important field during World War II and the Cold War that follows, culminating with the famous Nash Equilibrium. The objective of the games as a decision tool is to maximize some utility function for all decision makers under uncertainty. However this technique does not explicitly accommodate multiple criteria for selection of alternatives, therefore we will not consider it in this paper.
2. *Social Choice theory*. This approach represents voting mechanisms that allows the majority to express a choice. Social Choice theory considers votes of many individuals as the instrument for choosing a preferred candidate or alternative. The Theory of Social Choice was studies extensively with notable theories such as Arrow's Impossibility Theorem. This type of decision-making is based on the ranking of choices by the individual voters, while the scores that each decision maker gives to each criterion of each alternative are not considered explicitly. Therefore, this methodology is less suitable for multi-criteria decision making in which each criterion in each alternative is carefully weighted by the decision makers. The most well-known election procedures are Plurality Rule (Most Votes Count), Majority Rule (Pairwise Comparison), Borda Rule, and Approval Voting.
3. *Group decision using expert judgment*. This approach deals with integrating the preferences of several experts into a coherent group position. Within the Expert Judgment approach, there are two minor styles denoted as Team Decision and Group Decision. Both styles differ in the degree of disagreement that the experts are allowed to have while constructing the common decision. The essence of the group decision making can be summarized as follows: (1) there is a set of options and a set of individuals (decision makers) who provide their preferences over the set of options; (2) the problem is to find an option (or a set of options) that is best acceptable to the group of decision makers.

Based on those explanations above, our method utilizes group decision using expert judgment since it supports both multi-criteria and participatory decision making. In order to find out the best acceptable decision to all participants, we assign consensus level to indicate the degree of disagreement among decision makers, which will be discussed further below.

### 2.1 Consensus in Group Decision Making

Consensus is traditionally meant as a strict and unanimous agreement of all the experts regarding all possible alternatives. Ness and Hoffman define consensus in [20] as "*Consensus is a decision that has been reached when most members of the team agree on a clear option and the few who oppose it think they have had a reasonable opportunity to influence that choice. All team members agree to support the decision.*" The expression of concerns and conflicting ideas is considered desirable and important. The goal of consensus is not the selection of several options, but the development of one decision which is the best for the whole group. It is synthesis and evolution, not competition and attrition.

Hence, we can conclude that consensus decision making requires:

- Sufficient time to explore all the information and opinions.
- Strong facilitative leadership.
- Commitment and effort to develop an atmosphere of honesty and openness in the group.
- Willingness to contribute their views and discuss their reasons.
- Willingness to improve their knowledge and refine their decision.
- Willingness to confront and resolve controversy and conflict.
- Willingness to learn and listen from others

Yang (2010) conclude that consensus is thought to lead to higher quality decisions than single leader decision since it has tendency to influence others therefore it improves common understanding within group's member as well. However, in terms of the speed of decision making, single decision maker produces slightly faster result than consensus building.

## 2.2 Consensus Methods and Measurements

There are several individual preference aggregation methods to achieve consensus, such as: consensus ranking, distance function, goal programming, Multi-criteria Decision Analysis (MCDA), social behavior approach (i.e.: *social choice theory* dan *social judgement scheme*). Based on our literature study, we present those aggregation methods and consensus measurements as in table 1 below.

Table 1. Aggregation methods and Consensus Measurements in Group Decision Making

| Aggregation method | Decision Maker (DM) input | DM with weight | Consensus | Reference |
|---|---|---|---|---|
| OWA | Linguistic label | Y | [0, 1] | [1] |
| OWA linguistic quantifier | Preference ordering, | N | [0, 1] | [11] |
| Dynamic AHP | Pairwise comparison matrices | Y | Eigen value pairwise distances | [17] |
| ELECTRE TRI | Individual preferences weight | Y | Social Judgement Scheme (SJS) Consensus weight | [22] |
| TOPSIS with interval data | Prior information weight using Bayesian vector network | Y | Distance of each alternative from the positive and negative ideal solution | [30] |
| ELECTRE TRI | Individual preferences weight | Y | Disagreement exploration between group member | [18] |
| FAHP | Pairwise comparison matrices | Y | Get weights to experts then aggregate fuzzy number | [4] |
| AHP & FAST & Prometheus model | Pairwise comparison matrices | Y | Negotiation in value based decision | [28] |
| Fuzzy MCDM | Trapezoidal fuzzy number | Y | Satisfaction degree using TOPSIS method | [2] |
| AHP | Pairwise comparison using linguistic label | Y | Aggregating belief vector to calculate closeness coefficient | [9] |
| MAUT | Individual utility preferences | Y | Utility consensus value [0, 1] | [8] |
| AHP | Pairwise comparison matrices | Y | Consistency Consensus Matrix (CCM) using GCI | [19] |
| Collaborative multicriteria | Fuzzy triangular with linguistic label | Y | Fuzzy majority concept [0, 1] | [23] |
| agreement | | | | |
| FAHP-FGP | Pairwise comparison matrices | Y | Weight aggregation | [16] |
| AHP | Pairwise comparison matrices | Y | Preferential differences weight & rank with satisfactory index | [12] |
| Interval Evidential Reasoning | Numeric & interval judgement | N | [0, 1] | [29] |
| Hybrid distance-based ideal-seeking consensus ranking model | Individual ranking preferences | N | Distance between each ideal matrix and initial preference matrix | [24] |
| Consensus model (Euclidiean distance) | Individual preferences weight | Y | [0, 1] | [6] |

Generally from the table above, we conclude that consensus methods using preference aggregation can be divided into eight methods. Each method has been extended and or modified, i.e.:

1. OWA (*Ordered Weighted Averaging Operator*) and its extension: OWA was introduced by Yager. We find some extensions to it, such as: *Weighted* OWA (WOWA) [23], *Fuzzy Linguistic* OWA (FLOWA) [1].
2. MODM (*Goal programming*). *Goal Programming* (GP) was introduced by Chanrnes and Cooper in 1961. GP is a mathematical programming technique designed to handle conflicting objectives. GP can be used with other MCDA methods to decrease the weighting values [16].
3. MCDA. MCDA has been employed in many decision making cases. The most popular is AHP and its variant to achieve consensus ([17]; [4]; [28]; [9]; [19]; [16]; [12]). Other methods also have been commonly used such as: ELECTRE TRI ([22]; [18]), TOPSIS [30], and MAUT [8].
4. Aggregation method based on distance function to measure preference similarity or dissimilarity. This method usually has been employed together with other methods, such as: MCDA ([17]; [22]. The distance is measured from: (1) the preference's difference between decision maker for each criterion and or for each alternative; or (2) the weighting's difference between criteria and or alternative; or (3) the combination of both (1) and (2).
5. Aggregation method based on preference rank and preference interval. For example, Author in [12] utilize preference rank; while [29] employs numerical and *interval judgment*.

Related to consensus measurement, author in [3] mention two major categories of consensus measurements, i.e.:
1. *Hard consensus measurement. Hard consensus* has an interval [0,1]. It has been employed in ([1]; [11]; [8]; [30]) and generally has several methods to calculate a consensus degree, i.e.:
   - Count number of experts. The simplest consensus measure method is to count the number of experts within the group. Usually, the ratio of the number counted to the total group is taken as the consensus.
   - Distance. The method measures the distances between decision makers. The consensus is a function of the distance.
   - Similarity/Dissimilarity. Similar to the distance measure, similarities or dissimilarities between decision makers can be measured. Thus consensus is a function of the similarity/ dissimilarity, where consensus is the increasing function of similarity and decreasing function of dissimilarity.

- Order-Based. Based on the evaluations from experts, the preference orders of all alternatives from each expert can be calculated. By comparing the order difference from expert and the aggregated group, the consensus is then measured.
2. *Soft consensus measurement.* In this measurement, consensus is defined by *linguistic label*, such as: "*most*". It works well as *linguistic quantifier,* as implemented in [11].

## 3. THE PROPOSED CONSENSUS MODEL

Our proposed model employs hard consensus, by assigning consensus level within interval [1, 0]. The consensus level is determined at the early stage of decision making process. Based on the consensus level, we can define the maximum distance between decision makers. In our method, the leader of decision makers is selected by her reputation, named as Supra Decision Maker (SDM). French et al. in [7] mention that the existence of SDM in Group Decision Making (GDM) observes the entire elicitation and decision analysis process for each individual and altruistically uses this knowledge to construct a single decision analysis for the group. Hence, the choice is made according to the SDM's analysis. In our method, SDM has a role as central advisor who leads the preference similarity among other decision makers. The more similar the other's decision with the SDM's decision means the higher consensus level achievement. However, we do not address here the problem how to elect a SDM through trust and reputation mechanisms. This problem is addressed more detailed in our other papers which can be found in [26] and [27].

The consensus achievement proposed here is a combination of consensus model proposed by author in [6] and Social Judgment Scheme (SJS). SJS model is developed by Davis in 1996 [21]. The SJS model approximates the group judgment as a weighted average of the group members' initial judgments. Each member's initial judgment is weighted according to its relative closeness to the other members' judgments: the weight given to a particular member declines exponentially as the distance between his or her judgment and SDM' judgments becomes greater. In other words, members central (i.e. SDM) in terms of their judgmental preference are more influential in the group judgment process, whereas peripheral members (or deviant members) are less influential. Tindale et al. (2002) mention that SJS has been proven well in an empirical test as found in Davis et al. [5]; Hulbert et al. [13]; Ohtsubo et al. [21]; Rigopoulos et al. [22].

### 3.1 Model Description

We combine consensus model and SJS into our model. However, our model evaluate every DM's decision by weighting each alternative A with respect to criterion C. Based on the evaluations (comparisons) of each alternative between SDM and other DMs, the Euclidean-like 'distances' between decision makers are calculated. The generalized consensus level is defined as 1 minus the maximum distance between two members. In the model, a minimum consensus degree is required in advance. If the computed generalized consensus degree is smaller than the required one, a procedure of consensus reaching starts in which the SDM asks related decision maker to adjust his preferences; otherwise social function from SJS approach is employed. The decision of all Decision Makers (SDM and DM) then can be aggregated after each DM for each alternative has its own weighting value. The model formulation will be discussed later in sub section 3.1.2.

Our model also differs from other approaches, as it facilitates decision refinement through trust and reputation mechanism. We use trust and reputation mechanism to elect a Supra Decision Maker (SDM). A SDM acts as a leader and advisor to other decision makers, hence we might say that she has individual centrality role to the group. We do not address how trust and reputation mechanism works in this paper. For more detailed explanation about trust and reputation mechanism can be found in Tundjungsari et al. In [26] and [27]. Figure 1 below shows how decision makers act and cooperate with others to generate consensus and later decision in a group decision making.

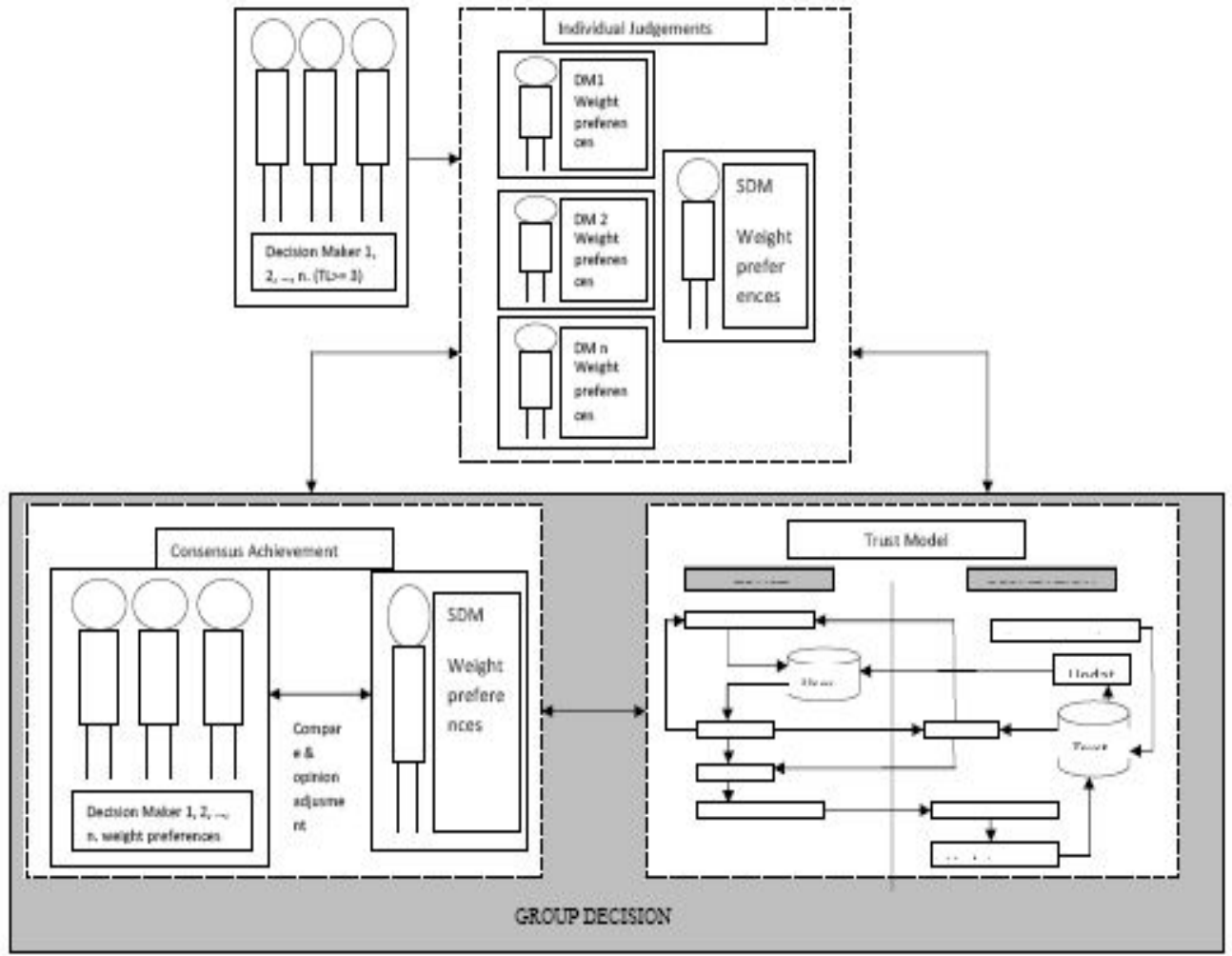

Figure 1. Schematic Illustration of Consensus Achievement Model

### 3.1.2 Formulation

Our model works as follows:
1) Let say that there are D decision makers (one of them is SDM). A is set of all alternatives and C is set of all criteria. Hence:
  Alternatives $A = \{a_i, i = 1, \ldots, m\}$
  Criteria $C = \{c_i, i = 1, \ldots, n\}$
  Decision Makers $D = \{d_i, i = 1, \ldots, n\}$
2) Every decision maker has to define his individual preferences by weigthing each criterium C, thus for each $i \in D$, we assume that $h_i : C \rightarrow [0,1]$. The preference of each alternative A in respect of each criterium C from each decision maker, is stated as $g_i(c, a)$. Hence for $i \in D$, define as $g_i: |C| \times |A|$ such that:

$$(f_i(a))_{a \in A} = (h_i(c))_{c \in C} \cdot (g_i(c,a))_{c \in C, a \in A}, \quad (1)$$

where $f_i(a)$ is i's evaluation of each alternative a for each criteria containing i's evaluation of each alternative.

3) The similarity between SDM and DM is calculated by Euclidian like distance as utilized in [6], hence d ($f_i$, $f_j$) where

    $f_i$ is DM's decision
    $f_j$ is SDM's decision

such that:

$$d(f_i, f_j) = \sqrt{\frac{1}{|A|} \sum_{a \in A} (f_i(a) - f_j(a))^2}. \quad (2)$$

The distance is used for determine whether a DM has achieve consensus level or not. If not, DM may change his preference. On the other hand, a SDM cannot change her preference since she has chosen by trust and reputation mechanisms, as the leader and advisor of the group for having the highest reputation value.

3) Consensus level θ is defined by the group in advance, such that:

    θ = 1 – max {d($f_i$, $f_j$) | i, j ∈ D}.     (3)

If d($f_i$, $f_j$) > max d($f_i$, $f_j$), a DM has to change his preferences until majority of his decision toward all alternatives has reached consensus level. For each decision maker, when half alternatives has reached consensus level, then we may say that consensus for decision maker i has been achieved. Hence, he may or may not change his decision toward other alternatives which has not fulfilled the required consensus level (for example, if the decision of DM i on 3 of 5 alternatives has attain consensus level then he may or may not change his decision toward other 2 alternatives). However, his weight preferences of alternatives which not within consensus degree will be decreased, as we employ social function w', such that:

    $w_i$' = exp [-θ(|$x_i$-$x_j$|)]     (4)

    where |$x_i$-$x_j$|= |d($f_i$, $f_j$) – max d($f_i$, $f_j$)|     (5)

As for each alternative that achieves minimum consensus level, the weighting w' automatically set to 1 (w'=1). The weighting value for every SDM's decision also set to 1.

5) All of the decision of each decision maker (SDM and DM) is aggregated toward its weighting value, such that:

$$f(a) = \sum_{i \in N^*} w'_i \cdot f_i(a), \quad (6)$$

By equation (6) above, we obtain the value of each alternative from all decision makers. Hence, we find out the rank of all alternatives A. Alternative with the highest value is assigned as first rank, and so on.

# 4. ILLUSTRATION

In this section, we present the example of our model. Suppose there is a group decision making (GDM) concerns to find out the rank of five community projects as available alternatives, such that: A = {a1, a2, a3, a4, a5}. There are three decision makers in group, which are: DM1, DM2, dan DM3, where DM3 is elected as a SDM, such that: D = {1, 2, 3}.

Each decision maker evaluates each alternative with respect to three criteria C = {c1, c2, c3}, where
- c1 is the project's urgency
- c2 is the project's impact to community
- c3 is the project's quality of the detailed work plan

Let say that GDM is agree on consensus level (θ) = 0.90. Therefore maximum distance allowed between DM and SDM is 0.1 (maximum distance = 1 – 0.90).

Suppose that each decision maker assigns criteria weighting value for three criteria available, such that:
- h1 = (0.7, 0.1, 0.1) ;
- h2 = (0.4, 0.1, 0.2) ;
- h3 = (0.3, 0.5, 0.2).

We assume that each criterium has four interval values to evaluate each alternative (i.e.: 1 is very high; 0.7 is high; 0.5 is moderate; 0.3 is low). Suppose that each decision maker evaluates alternatives with respect to criteria, as follows:

$$(g1\,(c,a))_{c \in C, a \in A} = \begin{pmatrix} 1 & 1 & 1 & 0.3 & 0.5 \\ 1 & 1 & 1 & 0.3 & 0.5 \\ 1 & 1 & 1 & 1 & 1 \end{pmatrix}$$

$$(g2\,(c,a))_{c \in C, a \in A} = \begin{pmatrix} 1 & 0.5 & 0.5 & 0.5 & 0.3 \\ 0.5 & 0.5 & 0.5 & 0.5 & 0.3 \\ 0.3 & 0.3 & 0.3 & 1 & 0.5 \end{pmatrix}$$

$$(g3\,(c,a))_{c \in C, a \in A} = \begin{pmatrix} 0.5 & 1 & 1 & 0.3 & 0.5 \\ 1 & 1 & 0.5 & 0.3 & 0.5 \\ 0.5 & 0.5 & 0.5 & 1 & 1 \end{pmatrix}$$

Thus, using equation (1) above, the decision makers' evaluation toward each alternative are:
(f1 (a)) $_{a \in A}$ = (0.9, 0.9, 0.9, 0.34, 0.5) ;
(f2 (a)) $_{a \in A}$ = (0.51, 0.31, 0.31, 0.45, 0.25) ;
(f3 (a)) $_{a \in A}$ = (0.75, 0.9, 0.65, 0.44, 0.6)

Using equation (2), we find out the distance between SDM and DM1, and the distance between SDM and DM2 in column [c] below. Column [d] is an evaluation toward distance maximum with respect to required consensus level, while column [e] is the weighting value as result to [4]. The weighting value is 1 (w'=1) when the decision value within consensus level; and w' < 1 (using equation 4) when the decision value is outside the required consensus level.

Table 2. Distances $d(f_i, f_j)$, where $i, j \in D = \{1, 2, 3\}$

| [a] Alternatives | [b] DM | [c] Distance (d) DM toward SDM (DM 3) | [d] d-dmax (Distance max = 0.1) | [e] Weighting value per alternative (w') |
|---|---|---|---|---|
| a1 | 1 (DM) | 0.15 | 0.0.5 | 0.951 |
|    | 2 (DM) | 0.24 | 0.14 | 0.869 |
| a2 | 1 (DM) | 0 | Consensus | 1 |
|    | 2 (DM) | 0.59 | 0.49 | 0.613 |
| a3 | 1 (DM) | 0.25 | 0.15 | 0.861 |
|    | 2 (DM) | 0.34 | 0.24 | 0.787 |
| a4 | 1 (DM) | 0.1 | Consensus | 1 |
|    | 2 (DM) | 0.01 | Consensus | 1 |
| a5 | 1 (DM) | 0.1 | Consensus | 1 |
|    | 2 (DM) | 0.35 | 0.25 | 0.779 |

From the table above, we can see that DM1 reaches consensus on 3 alternatives (a2, a4 and a5) from 5 available alternatives; while DM2 achieves consensus only on 1 alternative (a4) from 5 available alternatives. Therefore, DM2 has not reached minimum consensus; as the required is 3 alternatives (half of five available alternatives) for every decision maker should within consensus level.

Suppose that DM1 satisfy with his decision, so that he does not change his preference since his decisions on three alternatives (a2, a4, a5) within consensus level; therefore other alternative (a1, a3) will get decreasing weighting utilized social function. On the other hand, DM2 has to change his previous decision. Let say that DM2 refine his answer on criteria and alternative evaluation, such that:

h2 = (0.4, 0.4, 0.2) ;

$$(g2'(c, a))_{c \in C, a \in A} = \begin{pmatrix} 1 & 1 & 1 & 0.5 & 0.5 \\ 0.5 & 1 & 0.5 & 0.5 & 1 \\ 0.3 & 0.5 & 0.3 & 1 & 1 \end{pmatrix}$$

Hence by using equation (1), we get:

$(f2'(a))_{a \in A} = (0.66, 0.9, 0.66, 0.6, 0.8)$ ;

Table 3. Distances $d(f_i, f_j)$, where $i, j \in D = \{2, 3\}$

| [a] Alternatives | [b] DM | [c] Distance (d) DM toward SDM (DM 3) | [d] d-dmax (Distance max = 0.1) | [e] Weighting value per alternative (w') |
|---|---|---|---|---|
| a1 | 2 (DM) | 0.09 | Consensus | 1 |
| a2 | 2 (DM) | 0 | Consensus | 1 |
| a3 | 2 (DM) | 0.01 | Consensus | 1 |
| a4 | 2 (DM) | 0.16 | 0.06 | 0.942 |
| a5 | 2 (DM) | 0.2 | 0.1 | 0.905 |

From the new decision in the table above, we can find out that DM2 now has reached consensus on 3 alternatives (a1, a2, a3) from 5 available alternatives; while DM1 remains the same (a2, a4,

a5). All DM (DM1, DM2, and DM3) has agreed upon alternative a2, since all of their decision on alternative 2 is within required consensus level.

Having this, we can continue on preference aggregation using equation (6), result in as follows:

Table 4. Aggregation's Result toward all Alternatives

|  | DM1 | DM2 | DM3 | Total value | Rank |
|---|---|---|---|---|---|
| Alternative1 (a1) | 0.856 | 0.66 | 0.75 | 2.266 | 2 |
| Alternative2 (a2) | 0.9 | 0.9 | 0.9 | 2.7 | 1 |
| Alternative3 (a3) | 0.774 | 0.66 | 0.65 | 2.084 | 3 |
| Alternative4 (a4) | 0.34 | 0.565 | 0.44 | 1.345 | 5 |
| Alternative5 (a5) | 0.5 | 0.723 | 0.6 | 1.823 | 4 |

From the table above, we can see the alternatives' rank, i.e.:
Rank 1: Project 2 (a2)
Rank 2: Project 1 (a1)
Rank 3: Project 3 (a3)
Rank 4: Project 5 (a5)
Rank 5: Project 4 (a4)

## 5. CONCLUSIONS

In this paper we presented a brief overview of our work towards constructing a new consensus model, along with a simple example which was used in order to illustrate the model. Our literature studies show that there have not been many researches in group decision making utilize social approach, such as: trust and reputation, social judgment scheme. We integrate social approach within multi-criteria environment to build a consensus among decision makers. Moreover, we believe that this approach contributes to a better understanding of consensus achievement within a group decision making setting by considering each decision maker's evaluation. However, we believe that centrality approach produces better common understanding within group's member so that it can deliver better quality of decision. In the future, we will attempt to implement a web-based prototype as group decision support systems and deploy it in the real community.